\documentclass[final,numberedheadings,epsfig]{aipproc}
\layoutstyle{6x9}
\renewcommand\XFMaddressmarkstyle{\alph}

\newcommand{ \dn }{${\rm dN_{ch}/d\eta}$}
\newcommand{\Np }{${\rm \langle N_{part}\rangle}$}
\begin{document}
\vglue -3.0cm 
\title{Systematics of Global Observables in Cu+Cu and Au+Au Collisions at RHIC Energies} 
\author{Rachid NOUICER for the PHOBOS Collaboration\footnotemark}
{address={Chemistry Department, Brookhaven National Laboratory, Upton, NY 11973-5000, USA}
}
\begin{abstract} 
Charged particles produced in Cu+Cu collisions at ${\rm \sqrt{s_{_{NN} }} = }$
200 and 62.4 GeV have been measured in the PHOBOS experiment at
RHIC. The comparison of the results for Cu+Cu and Au+Au for the most
central collisions at the same energy reveals that the particle
density per nucleon participant pair and the extended longitudinal
scaling behavior are similar in both systems. This implies that 
for the most central events in symmetric nucleus-nucleus collisions
the particle density per nucleon participant pair does not depend on
the size of the two colliding nuclei but only on the collision energy.
Also the extended longitudinal scaling seems independent of the
colliding energy and species for central collisions. In addition,
there is an overall factorization of \dn\ shapes as a function of
collision centrality between Au+Au and Cu+Cu collisions at the same
energy.
\end{abstract}
\maketitle \footnotetext{Collaboration: B.Alver$^4$, B.B.Back$^1$,
M.D.Baker$^2$, M.Ballintijn$^4$, D.S.Barton$^2$, R.R.Betts$^6$,
R.Bindel$^7$, W.Busza$^4$, Z.Chai$^2$, V.Chetluru$^6$,
E.Garc\'{\i}a$^6$, T.Gburek$^3$, K.Gulbrandsen$^4$, J.Hamblen$^8$,
I.Harnarine$^6$, C.Henderson$^4$, D.J.Hofman$^6$, R.S.Hollis$^6$,
R.Ho\l y\'{n}ski$^3$, B.Holzman$^2$, A.Iordanova$^6$, J.L.Kane$^4$,
P.Kulinich$^4$, C.M.Kuo$^5$, W.Li$^4$, W.T.Lin$^5$, C.Loizides$^4$,
S.Manly$^8$, A.C.Mignerey$^7$, R.Nouicer$^2$, A.Olszewski$^3$,
R.Pak$^2$, C.Reed$^4$, E.Richardson$^7$, C.Roland$^4$, G.Roland$^4$,
J.Sagerer$^6$, I.Sedykh$^2$, C.E.Smith$^6$, M.A.Stankiewicz$^2$,
P.Steinberg$^2$, G.S.F.Stephans$^4$, A.Sukhanov$^2$, A.Szostak$^2$,
M.B.Tonjes$^7$, A.Trzupek$^3$, G.J.van~Nieuwenhuizen$^4$,
S.S.Vaurynovich$^4$, R.Verdier$^4$, G.I.Veres$^4$, P.Walters$^8$,
E.Wenger$^4$, D.Willhelm$^7$, F.L.H.Wolfs$^8$, B.Wosiek$^3$,
K.Wo\'{z}niak$^3$, S.Wyngaardt$^2$, B.Wys\l ouch$^4$. $^1$~Argonne
National Laboratory, Argonne, IL 60439-4843, USA.  $^2$~Brookhaven
National Laboratory, Upton, NY 11973-5000, USA.  $^3$~Institute of
Nuclear Physics PAN, Krak\'{o}w, Poland.  $^4$~Massachusetts Institute
of Technology, Cambridge, MA 02139-4307, USA. $^5$~National Central
University, Chung-Li, Taiwan.  $^6$~University of Illinois at Chicago,
IL 60607-7059, USA.  $^7$~University of Maryland, MD 20742,~USA.
$^8$~University of Rochester, Rochester, NY 14627,~USA.}
\vspace*{-1.2cm}
\section*{Introduction}
\vspace*{-0.2cm}
In relativistic heavy ion physics, it is well established that the
global observables and their spatial distributions are indicators for
the global reaction dynamics and kinematics. They are very useful
tools for event characterization. But one can also study their
systematics as a function of beam energy and system size which may
shed light on the onset of the formation of a new state of
deconfinement. Recently, the multiplicity of charged particles
produced in Cu+Cu collisions at ${\rm \sqrt{s_{_{NN} }} = }$ 200 and
62.4 GeV has been measured with the PHOBOS detector at RHIC.  These
measurements provide an excellent opportunity to study the systematics
of charged particle multiplicity by comparing the results obtained in
Cu+Cu and Au+Au collisions of the same energies.\par The Cu+Cu data
were collected using a silicon multiplicity array, which consists of
an octagonal multiplicity detector and six forward silicon counters,
three on each side of the interaction point. Two analysis methods: a
``hit-counting'' method and an ``analog'' method were
used~\cite{BacPRL2001}. The measured~\dn\ was corrected for particles
which were absorbed or produced in the surrounding material and for
feed-down products from weak decays of neutral strange particles.  The
centrality determination was obtained by using charged particles
detected in two sets of 16 scintillators counters located at ${z =
\pm}$ 3.21 meters from the nominal interaction point along the beam
axis. The multiplicity array, the analysis procedures and the
centrality determination used for Cu+Cu and Au+Au collisions are the
same.
\begin{center}
\begin{figure}[btp]
\centering
  \begin{tabular}{cc}
    \begin{minipage}{3in}
      \centering
      \includegraphics[height=.27\textheight]
		      {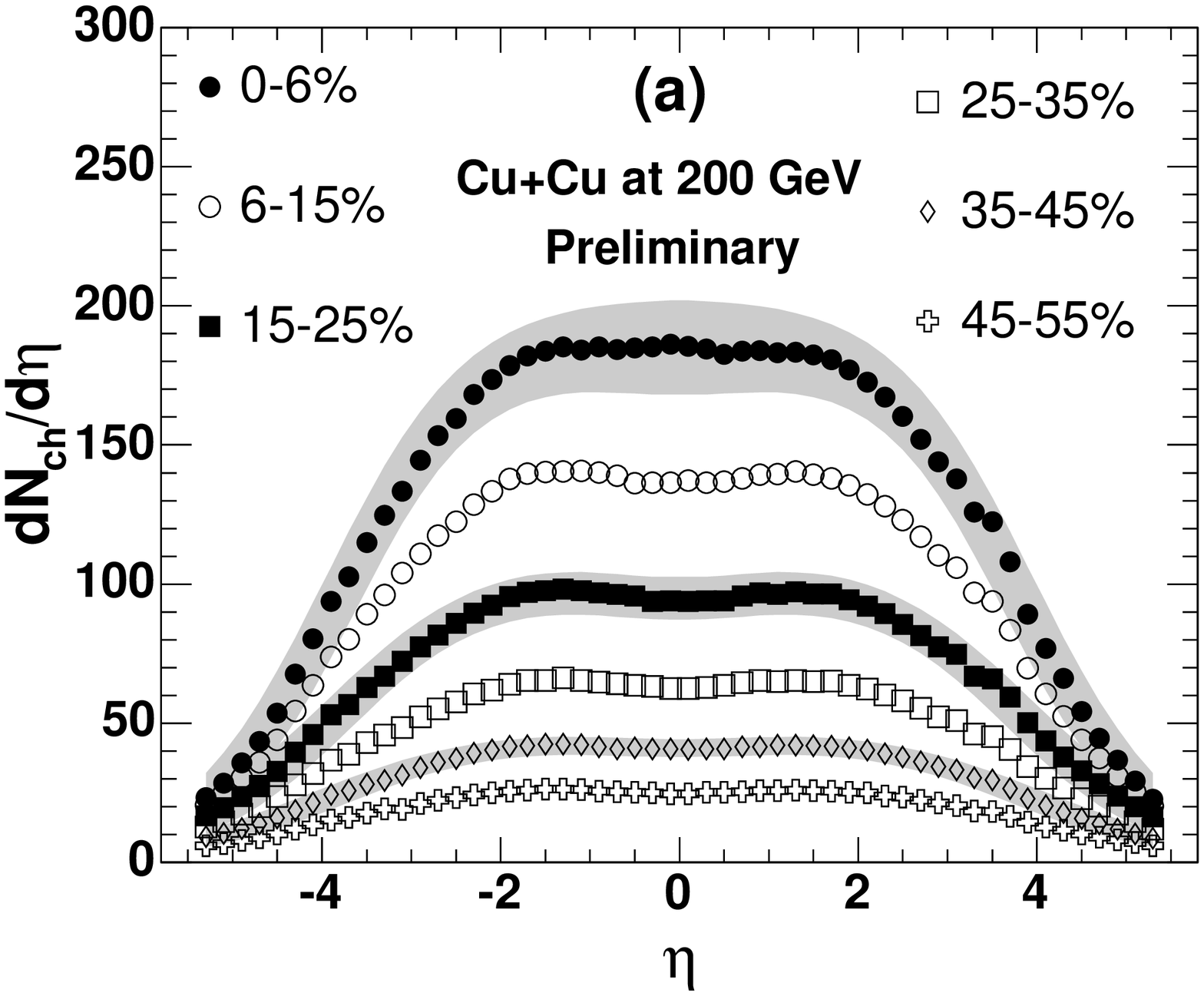}
    \end{minipage}
    &
    \hspace*{-0.5cm}\begin{minipage}{3in}
      \centering
      \includegraphics[height=.27\textheight]
		      {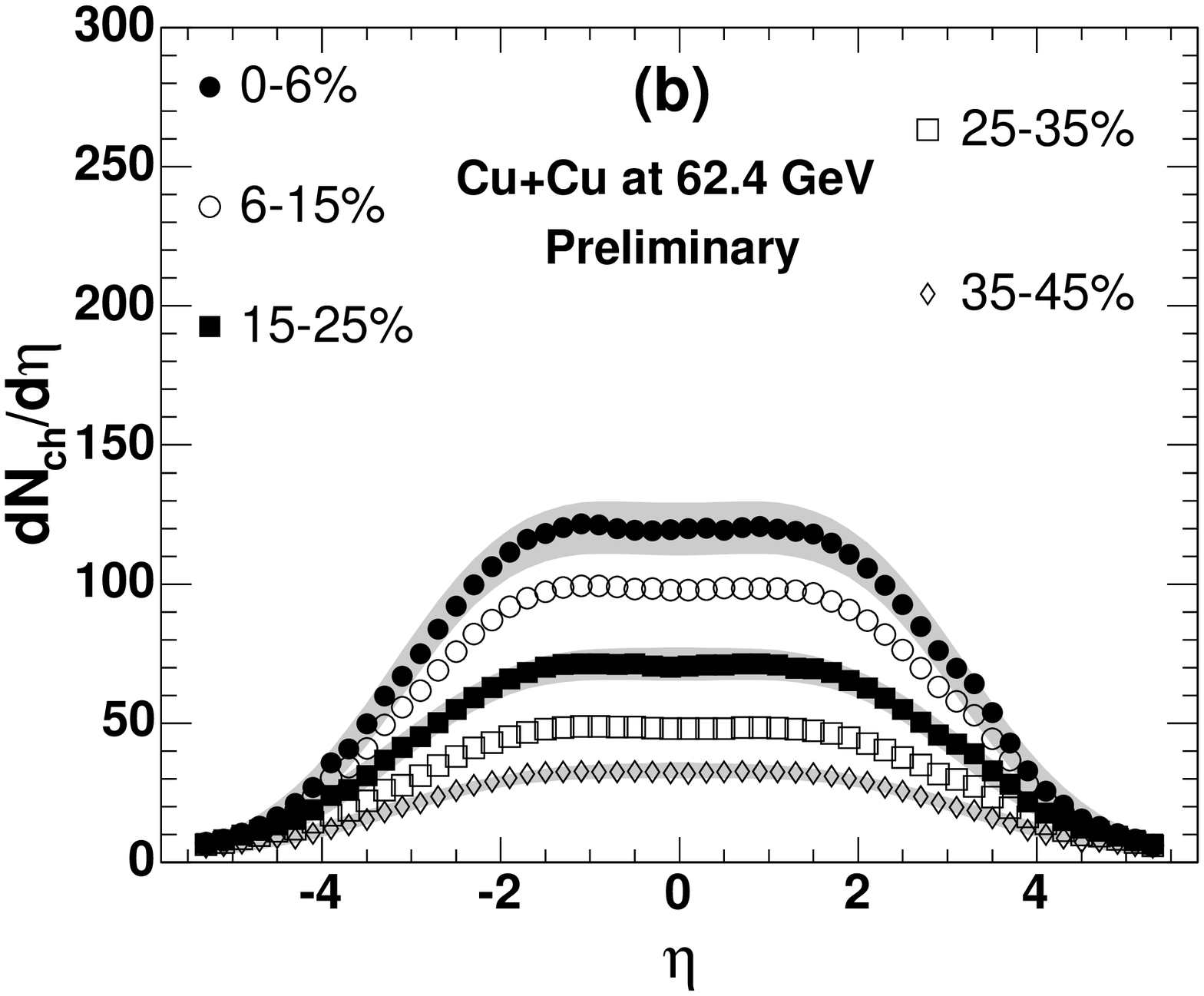}
    \end{minipage}
    \\
    \begin{minipage}{3in}
      \centering
      \includegraphics[height=.27\textheight]
		      {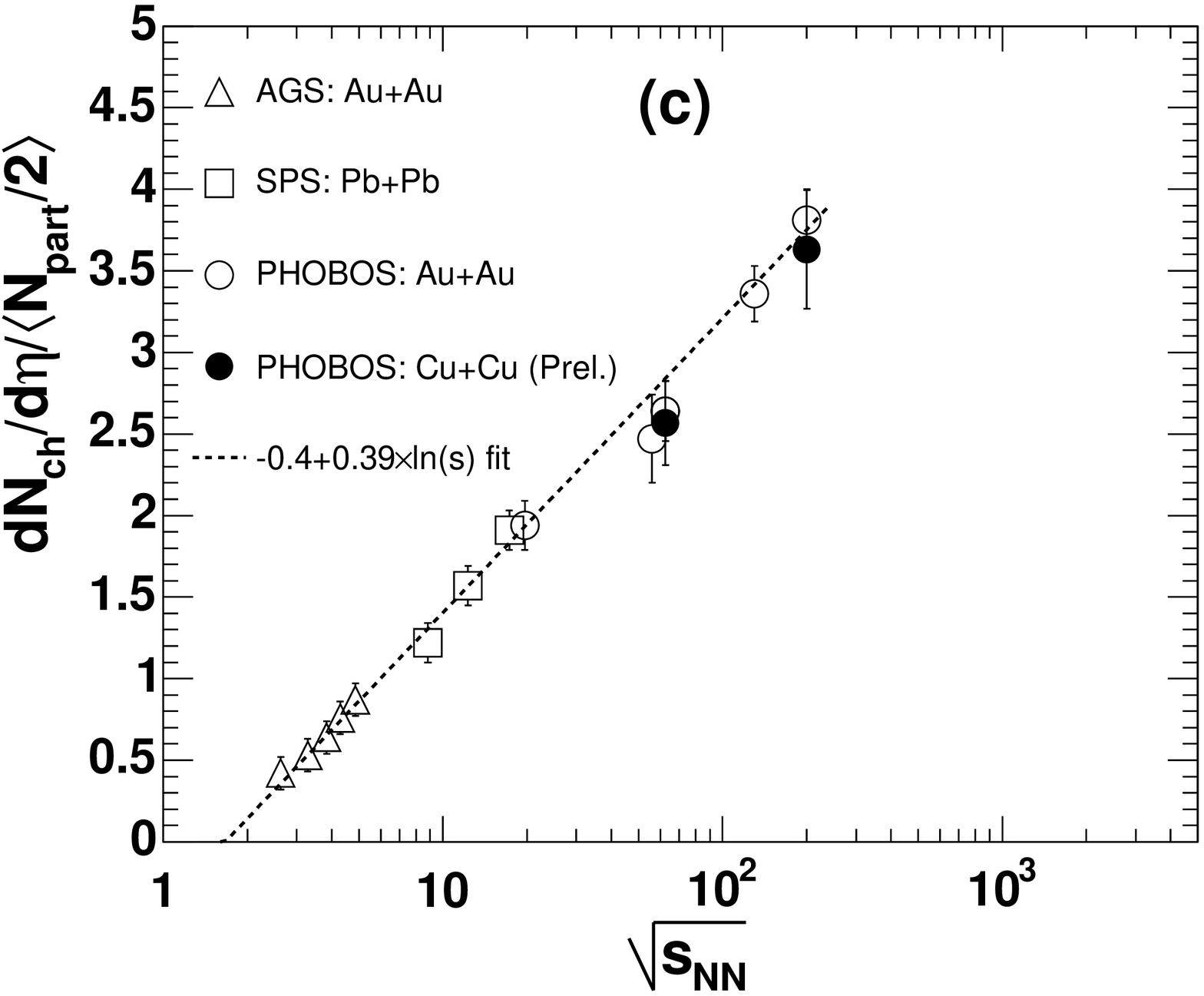}
    \end{minipage}
    &
    \begin{minipage}{3in}
      \centering
      \hspace*{-0.5cm}\includegraphics[height=.27\textheight]
	      {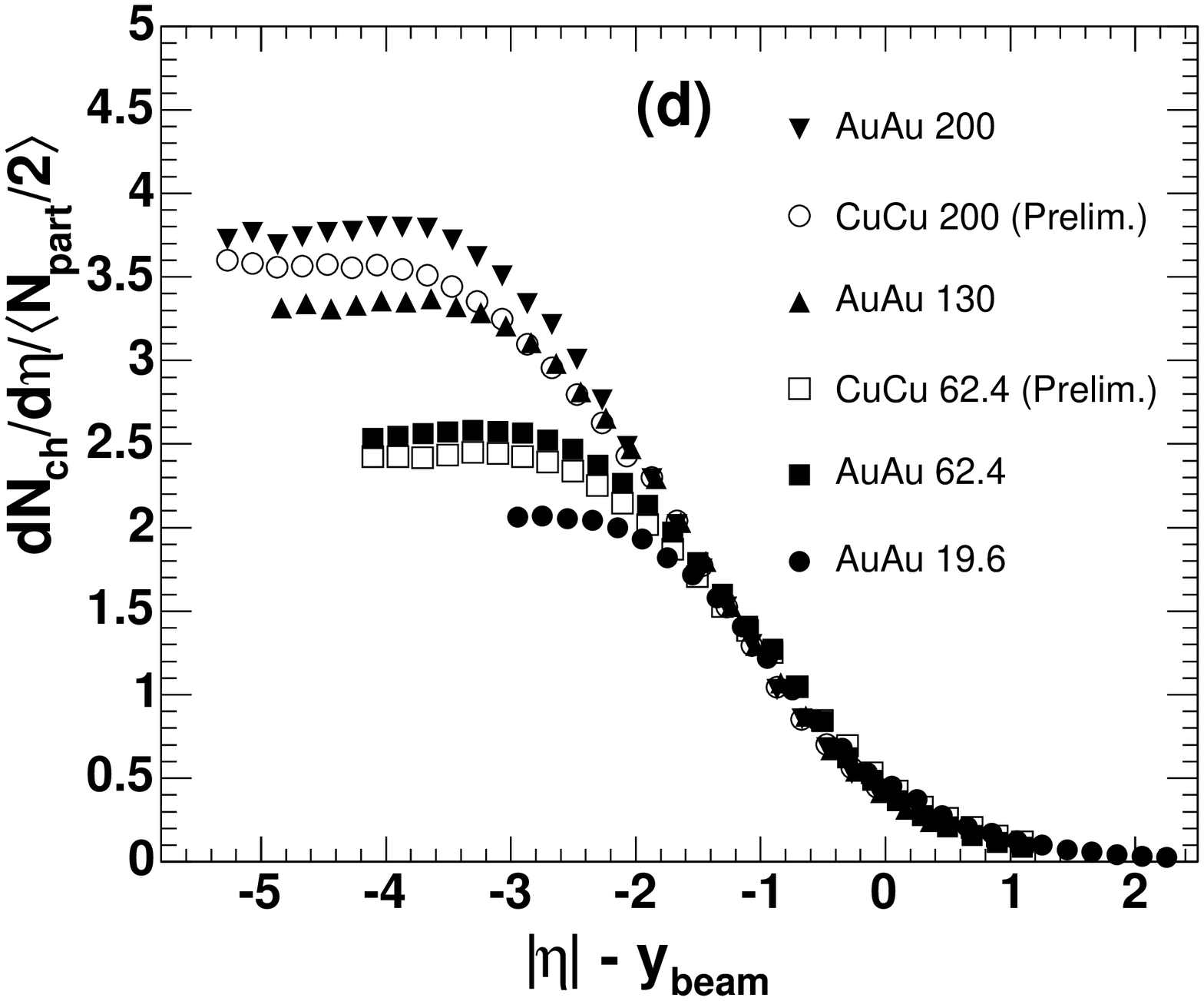}
    \end{minipage}
  \end{tabular}
  \caption{Panels a) and b): Measured~\dn~distributions of charged
  particles from Cu+Cu collisions at 200 and 62.4 GeV as a function of
  collision centrality (preliminary results). Shaded bands represent
  the systematic errors. Panel c): Particle density per nucleon
  participant pair produced in central (6\%) nucleus-nucleus
  collisions as a function of energy from AGS, SPS and
  RHIC data~\cite{RN2005,Back2005}. Panel d): Pseudorapidity distributions
  for Au+Au and Cu+Cu central collisions (6\%) at several RHIC
  energies~\cite{RN2005,Gunt}. The distributions have been shifted to
  ${\rm \mid\eta\mid -y_{beam}}$ in order to effectively study the
  fragmentation regions in the rest frame of one nucleus.}
  \label{fig1}
\end{figure}
\end{center}
\vspace*{-2.2cm}
\section*{RESULTS AND CONCLUSIONS}
\vspace*{-0.2cm}
The~\dn\ distributions measured for Cu+Cu collisions at 200 and 62.4
GeV are presented in Figures~\ref{fig1}.a) and b), respectively.  We
observe that there is a smooth transition between the mid-rapidity
plateau and the fragmentation region~\cite{RN2005}.
Figure~\ref{fig1}.c) shows the charged particle density per nucleon
participant pair in the mid-rapidity region (${\rm |\eta| \le 1}$) for A+A collisions from
AGS to RHIC energies. The comparison of the results for Au+Au and
Cu+Cu most central collisions indicate, that for the most
central events in symmetric nucleus-nucleus collisions the density per
nucleon participant pair does not depend on the size of the two
colliding nuclei but only on the collision energy~\cite{RN2005}. In
the fragmentation region, the comparison of~\dn~distributions per
nucleon participants pair produced in Cu+Cu and Au+Au collisions at
several energies are presented in Figure~\ref{fig1}.d). We observe
that the Cu+Cu and Au+Au collisions exhibit the same extended
longitudinal scaling for central collisions~\cite{Gunt}.\par

\begin{figure}[t]
  \includegraphics[height=.25\textheight]{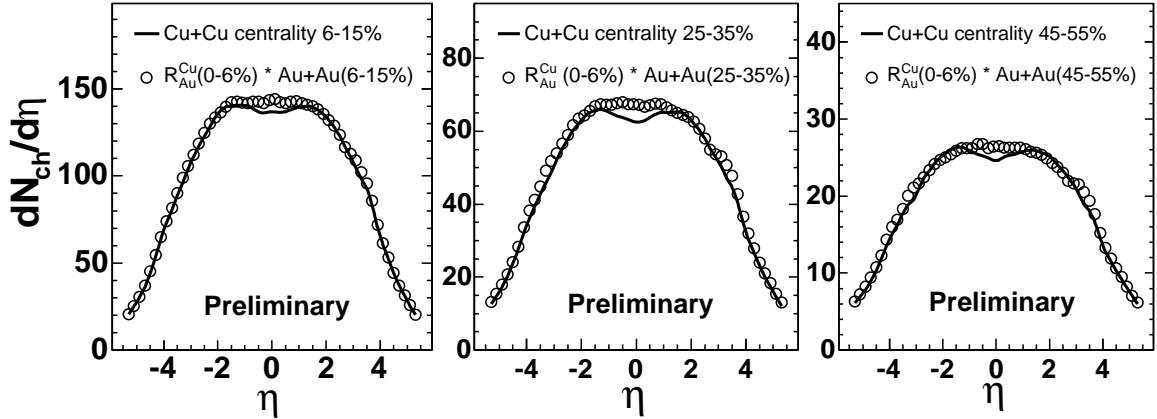}
\vspace*{-0.6cm}\caption{Comparison of \dn\ distributions for Cu+Cu
and Au+Au collisions at the same centrality and energy (200 GeV),
presented for different centrality bins~\cite{RN2005}. The \dn\
distributions of Au+Au collisions have been multiplied by the ratio of
the measured \dn\ distributions of Cu+Cu central (6\%) collisions to
the measured \dn\ of Au+Au central (6\%) collisions. For clarity, the
systematic errors are not shown.}
\label{fig2}
\end{figure}

For 0-6\% central collisions we find that the multiplicity shapes are
essentially identical for Au+Au and Cu+Cu collisions, differing only
by a factor which is approximately ${\rm
\eta}$-independent~\cite{RN2005}. The ratio of\ \dn\ distributions
measured for the two systems (Cu+Cu to Au+Au) for 6\% most central
collisions was used to scale the\ \dn\ distributions measured in Au+Au
collisions with different centralities. Figure~\ref{fig2} shows
that~\dn\ for Cu+Cu collisions and scaled~\dn\ for Au+Au system are
very similar, indicating that the shapes of pseudorapidity
distributions are the same for both systems and all centrality bins~\cite{RN2005}.
The small difference at mid-rapidity can be related to the difference
of the mean $P_{T}$ of charged particle in Cu+Cu and Au+Au collisions
but it falls well within the systematic errors. It thus appears that
the ~\dn~shapes are independent of the overall size of the colliding
nuclei at least between the Cu+Cu and Au+Au systems studied here.

\par In this talk, we have shown that the particle density and
extended longitudinal scaling of charged particles are similar in
Cu+Cu and Au+Au for central collisions at the same energy. In
Ref.~\cite{Gunt} we have shown that the\ \dn\ distributions are
identical for ${\rm |\eta|\le 4}$ for Cu+Cu and Au+Au collisions
characterized by the same\ \Np. In the present work we observe
that at 200 GeV the shapes of\ \dn\ distributions are similar for
the two systems in the full ${\rm |\eta|}$ range for collisions in the same
centrality class as defined by the fraction of the total inelastic
cross section.
\par
\vspace*{0.1cm}
\noindent{
\footnotesize	
This work was partially supported by U.S. DOE grants
DE-AC02-98CH10886,
DE-FG02-93ER40802,
DE-FC02-94ER40818,  
DE-FG02-94ER40865,
DE-FG02-99ER41099, and
W-31-109-ENG-38, by U.S.
NSF grants 9603486, 
0072204,            
and 0245011,        
by Polish KBN grant 1-P03B-062-27(2004-2007),
by NSC of Taiwan Contract NSC 89-2112-M-008-024
and by Hungarian OTKA grant (F 049823).
}


\vspace*{-0.9cm}
\bibliographystyle{aipprocl} 
\bibliography{sample}
\IfFileExists{\jobname.bbl}{}
 {\typeout{}
  \typeout{******************************************}
  \typeout{** Please run "bibtex \jobname" to optain}
  \typeout{** the bibliography and then re-run LaTeX}!
  \typeout{** twice to fix the references!}
  \typeout{******************************************}
  \typeout{}
 }

\end{document}